%% file: main.tex
\documentclass[12pt]{iopart}

\usepackage{iopams}  
\usepackage{adjustbox}
\usepackage{hyperref}
\usepackage[numbers,sort&compress,square]{natbib}
\usepackage{tabularx}
\usepackage{xcolor}
\usepackage[normalem]{ulem}

\begin{document}

\title[Multi-view Fusion with Attention-based Machine Learning for O4]{Advancing Glitch Classification in Gravity Spy: Multi-view Fusion with Attention-based Machine Learning for Advanced LIGO's Fourth Observing Run}

\author{Yunan~Wu$^1$, Michael~Zevin$^{2,3}$, Christopher~P.~L.~Berry$^4$, Kevin~Crowston$^5$, Carsten~{\O}sterlund$^5$, Zoheyr~Doctor$^{3,6}$, Sharan~Banagiri$^3$, Corey~B.~Jackson$^7$, Vicky~Kalogera$^{3,6}$, Aggelos~K.~Katsaggelos$^{1,3}$}

\address{$^1$ The Department of Electrical Computer Engineering, Northwestern University, 2145 Sheridan Road, Evanston, 60208, IL, USA}
\address{$^2$ The Adler Planetarium, 1300 South DuSable Lake Shore Drive, Chicago, 60605, IL, USA}
\address{$^3$ Center for Interdisciplinary Exploration and Research in Astrophysics (CIERA), Northwestern University, 1800 Sherman Ave, Evanston, 60201, IL, USA} 
\address{$^4$ SUPA, School of Physics and Astronomy, University of Glasgow, Kelvin Building, University Ave, Glasgow, 8QQ, G12, UK}
\address{$^5$ School of Information Studies, Syracuse University, Hinds Hall, Syracuse, 13210, NY, USA}
\address{$^6$ Department of Physics and Astronomy, Northwestern University, 2145 Sheridan Road, Evanston, 60208, IL, USA}
\address{$^7$ Information School, University of Wisconsin–Madison, 600 N Park Street, Madison, 53706, WI, USA}

\ead{christopher.berry.2@glasgow.ac.uk}

\begin{abstract}
The first successful detection of gravitational waves by ground-based observatories, such as the Laser Interferometer Gravitational-Wave Observatory (LIGO), marked a breakthrough in our comprehension of the Universe. 
However, due to the unprecedented sensitivity required to make such observations, gravitational-wave detectors also capture disruptive noise sources called glitches, which can potentially be confused for or mask gravitational-wave signals. 
To address this problem, a community-science project, Gravity Spy, incorporates human insight and machine learning to classify glitches in LIGO data. 
The machine-learning classifier, integrated into the project since 2017, has evolved over time to accommodate increasing numbers of glitch classes. 
Despite its success, limitations have arisen in the ongoing LIGO fourth observing run (O4) due to the architecture's simplicity, which led to poor generalization and inability to handle multi-time window inputs effectively. 
We propose an advanced classifier for O4 glitches. 
Using data from previous observing runs, we evaluate different fusion strategies for multi-time window inputs, using label smoothing to counter noisy labels, and enhancing interpretability through attention module-generated weights. 
Our new O4 classifier shows improved performance, and will enhance glitch classification, aiding in the ongoing exploration of gravitational-wave phenomena.
\end{abstract}

%
\submitto{\CQG}
%
%
%

\input{Introduction/intro}

\input{Materials_and_Methods/Materials}

\input{Results/results}

\input{Discussion/discussion}

\input{Conclusion/conclusion}

\section*{Acknowledgments}
The authors extend their gratitude to the community-science volunteers of the Gravity Spy project, whose dedicated efforts played a pivotal role in bringing this  project to fruition. 
The authors also thank ManLeong Chan and the anonymous referees for comments that improved the quality of this manuscript, and Shamal Lalvani, Yi Li and Jonathan Stromer-Galley for insightful conversations. 
This material is based upon work supported by NSF's LIGO Laboratory which is a major facility fully funded by the National Science Foundation. 
The authors are grateful for computational resources provided by the LIGO Laboratory and supported by National Science Foundation Grants PHY-0757058 and PHY-0823459. Gravity Spy is supported by NSF grants IIS-2106882, 2106896, 2107334, and 2106865. 
CPLB acknowledges support from Science and Technology Facilities Council (STFC) grant ST/V005634/1. 
ZD acknowledges support from the CIERA Board of Visitors Research Professorship and NSF grant PHY-2207945. 
This document has been assigned LIGO document number LIGO-P2300458.

\section*{Declaration of Competing Interest}
The authors declare that they have no known competing financial interests or personal relationships that could have appeared to influence the work reported in this paper.

\section*{Data Availability Statement}
The datasets from the Gravity Spy project, including machine learning and volunteer classification information, can be accessed through the Zenodo repositories \cite{glanzer2021gravity, zevin2022gravity}. 
The corresponding codes are also accessible on GitHub through \href{https://github.com/Gravity-Spy/gravityspy-ligo-pipeline}{Gravity Spy}.

\bibliographystyle{iopart-num}
\bibliography{bib}

\end{document}

%% file: Introduction/intro.tex
\section{Introduction}

The first discovery of gravitational waves, a pivotal element in Einstein's theory of general relativity~\cite{einstein1916approximative}, has opened up an entirely new window in the cosmos. The Laser Interferometer Gravitational-Wave Observatory (LIGO)~\cite{aasi2015advanced} achieved groundbreaking success in detecting these ripples in spacetime for the first time in September 2015~\cite{abbott2016observation}. 
Since then, the gravitational-wave detector network has been expanded with the addition of Virgo~\cite{acernese32advanced}, which joined observations in August 2017~\cite{PhysRevLett.119.141101, abbott2019gwtc}, and KAGRA~\cite{kagra2019kagra}, which initiated its observing run in April 2020~\cite{ligo2022first}; this network has collected a large catalog of gravitational-wave observations~\cite{abbott2023gwtc3}.
However, collecting these observations demands exceptionally sensitive and intricate detectors in order to be able to measure minuscule fluctuations in spacetime~\cite{abbott2020guide}. 
This heightened sensitivity, in turn, leads to the detection of diverse sources of noise, with the potential to obscure or mimic authentic gravitational-wave signals~\cite{abbott2016characterization, davis2021ligo, abbott2023gwtc3, berger2023searching}. 
Bursts of non-Gaussian noise caused by environmental or instrumental factors, known as glitches, are particularly disruptive to measuring gravitational waves. 
The origins of many glitches remain largely unknown~\cite{davis2022detector, nuttall2018characterizing, cabero2019blip}. 
The process of characterizing and eliminating these glitches is essential to enhancing the quality of the detection system and the analysis of detector data.

In pursuit of a better understanding of detector noise, \emph{Gravity Spy}~\cite{zevin2017gravity, zevin2023gravity}, a community-science project, has been launched to leverage the power of both volunteers and machine learning to classify glitches in LIGO data. 
On one front, the project engages human expertise in identifying and categorizing various glitches collected by LIGO. 
These labeled glitches then serve as training data to refine machine-learning classifiers, enhancing the accuracy and efficiency of glitch identification. 
Conversely, the classifiers provide guidance to volunteers, assisting them in identifying existing or new potential glitch classes~\cite{coughlin2019classifying}. 
Gravity Spy is implemented on the Zooniverse platform~\cite{fortson2018optimizing} on the community-science side, and it has demonstrated significant success by attracting over 30 thousand volunteers making more than 7 million glitch classifications.\footnote{\href{https://gravityspy.org}{gravityspy.org}} 
The outputs are actively used by LIGO detector-characterization experts~\cite{davis2021ligo, soni2024ligo}. 
One critical component contributing to the results of Gravity Spy is the integration of the machine-learning classifier.

The machine-learning classifier has been seamlessly integrated into the Gravity Spy pipeline for over five years, largely maintaining its original architecture. 
Periodically, fine-tuning of its final layer has been performed to enhance its adaptability to diversely predicted glitch classes~\cite{dodge2020fine}. 
The input for the classifier remains unchanged, featuring a structure that juxtaposes four spectrograms of the data from different time windows ($0.5~\mathrm{s}$, $1.0~\mathrm{s}$, $2.0~\mathrm{s}$ and $4.0~\mathrm{s}$, as shown in Figure~\ref{fig:glitch_ex}) of a single glitch into one merged image. 
The same images are shown to volunteers on Zooniverse for their classifications~\cite{zevin2017gravity}. 

The first version of the machine-learning classifier was introduced in the initial Gravity Spy paper~\cite{zevin2017gravity} and was integrated into the Gravity Spy pipeline after the completion of the first observing run (O1). 
Designed for a 20-class classification setup, this classifier consisted of two convolutional layers and two fully-connected layers to extract image features. 
Each convolutional layer was followed by a max pooling layer to reduce the dimensionality of the features. 

As the project progressed, during the second observing run (O2), Gravity Spy volunteers identified new glitch classes and two of them, 1080 Lines and 1400 Ripples, were added into the classifier. Therefore, the classifier was optimized by incorporating two additional convolutional layers and max pooling layers before the existing fully-connected layers, which facilitated the extraction of more discriminative features and adapted the classifier to a $22$-class classification task~\cite{bahaadini2018machine}. 

During the third observing run (O3), two new glitch classes, known as Low-frequency Blip and Fast Scattering (or Crown), were identified through the collaborative efforts of volunteers and LIGO detector-characterization experts~\cite{soni2021discovering}. 
Alongside this, the decision was made to remove the None of the Above class from the training dataset. 
The None of the Above class was initially aimed at empowering our volunteers to identify new glitch classes. 
However, its inclusion confuses the model, especially given morphological similarities with other glitches. 
As a result, removing this class from the model allows the identification of new glitches more effectively with low predicted confidence~\cite{soni2021discovering, glanzer2023data}. 
This led to further retraining of the classifier, enabling it to effectively accommodate a 23-class classification task~\cite{soni2021discovering}. 
    
Currently, the fourth observing run (O4) of the gravitational-wave detector network is in progress~\cite{abbott2020prospects, abac2024observation, capote2024advanced}. 
With the anticipation of discovering new glitch classes during O4 and subsequent runs, the current Gravity Spy classifier faces major limitations~\cite{alvarez2023gspynettree, jarov2023new, glanzer2023data}. 
First, in the glitch classification task, unlike other image processing tasks, location plays a crucial role. 
For example, distinguishing between a line at the top and a line at the bottom of the image is essential as they represent different glitches at different frequencies. 
However, the classifier's relatively straightforward architecture, comprised mainly of a combination of convolutional and fully-connected layers, hinders its capacity to capture unique features in individual glitch classes \cite{simonyan2014very}. 
Second, due to its shallow layer structure, the classifier requires significant resizing of input image dimensions, resulting in the loss of valuable information. 
Moreover, previous studies~\cite{bahaadini2017deep} have shown that the classification accuracy of glitches varies depending on the time window used. 
However, earlier classifier studies~\cite{zevin2017gravity, bahaadini2017deep} relied on a single model to extract features from merged image inputs across multiple time windows, limiting their ability to capture cross-time window correlations~\cite{huang2020fusion}. 
Recognizing these challenges, it becomes apparent that the demand to analyze more data and diverse glitch types exceeds the current model's capabilities. 

Additionally, the current classifier grapples with a confidence overfitting issue~\cite{guo2017calibration, nguyen2015deep}, displaying excessive certainty in glitch predictions even when errors are present. 
This tendency can potentially mislead both volunteers and LIGO detector-characterization experts, posing a risk to the efficiency and reliability of glitch characterization. Therefore, there is a compelling need for an advanced classifier architecture with robust generalization capabilities in the context of O4.

In this study, we develop a novel machine-learning classifier for glitch classification in the ongoing O4 run to address the challenges faced by Gravity Spy. 
The proposed classifier employs an attention-based multi-view fusion strategy to capture cross-time window correlations across the four time windows. 
The incorporation of a regularizer and label smoothing techniques into the loss function improves generalization and mitigates the confidence overfitting issue. 
The contributions of this paper can be summarized as follows: (1) we comprehensively compare three fusion strategies (early fusion, late fusion, and intermediate fusion) in scenarios with multiple time windows as inputs; 
(2) we apply label smoothing to mitigate the impact of noisy training labels, 
and (3) we introduce attention modules in the classifier, enhancing transparency in the glitch decision-making process by identifying specific window time that predominantly influences the glitch classification decision---an exploration undertaken for the first time. 

The paper is organized as follows: 
In Section~\ref{'sec:2-methods'}, we provide an overview of the Gravity Spy dataset, introduce three fusion strategies, outline the architecture of the proposed classifier, and elaborate on the conducted statistical analysis. 
Moving on to Section~\ref{'sec:3-results'}, we present the outcomes obtained from various ablation studies. 
In Section~\ref{'sec:4-discussion'}, we delve into a comprehensive discussion, addressing both the strengths and limitations of this study. 
We conclude in Section~\ref{'sec:5-conclusion'}.

%% file: Materials_and_Methods/Materials.tex
\section{Materials and methods}
\label{'sec:2-methods'}

In this section, we provide an overview of the Gravity Spy dataset, covering its data-generation process, image preprocessing, and the methodology for data splitting during both training and testing phases. 
Following this, we delve into the multi-view fusion strategy, discussing its early, intermediate, and late fusion components. 
Last, we propose a novel classifier, outlining the model architecture, attention modules, and the application of a label-smoothing technique.

\subsection{Gravity Spy dataset}\label{'sec:2-methods-dataset'}

The Gravity Spy dataset comprises time--frequency spectrograms of glitches.

The glitches are initially identified by the Omicron pipeline~\cite{robinet2016omicron}. 
Omicron searches for excess power in the data by performing a constant-$Q$ transform~\cite{brown1991calculation}, which projects the data onto a basis of Gaussian-windowed complex-valued exponential functions that tile the time--frequency plane ($Q$ refers to the ratio of duration to bandwidth), and then searching for significant clusters of tiles. 
Omicron calculates signal-to-noise ratios (SNRs) for triggers from the magnitudes of the most significant time--frequency tiles~\cite{chatterji2005search}.  
To concentrate on glitches that pose significant challenges to gravitational-wave analyses (and are visible in spectrograms), the Gravity Spy pipeline only considers triggers with an Omicron  SNR greater than $7.5$, and where the peak frequency falls between  $10~\mathrm{Hz}$ and $2048~\mathrm{Hz}$.   

\begin{figure} 
\begin{center}
\begin{tabular}{c} 
\includegraphics[width=0.95\textwidth]{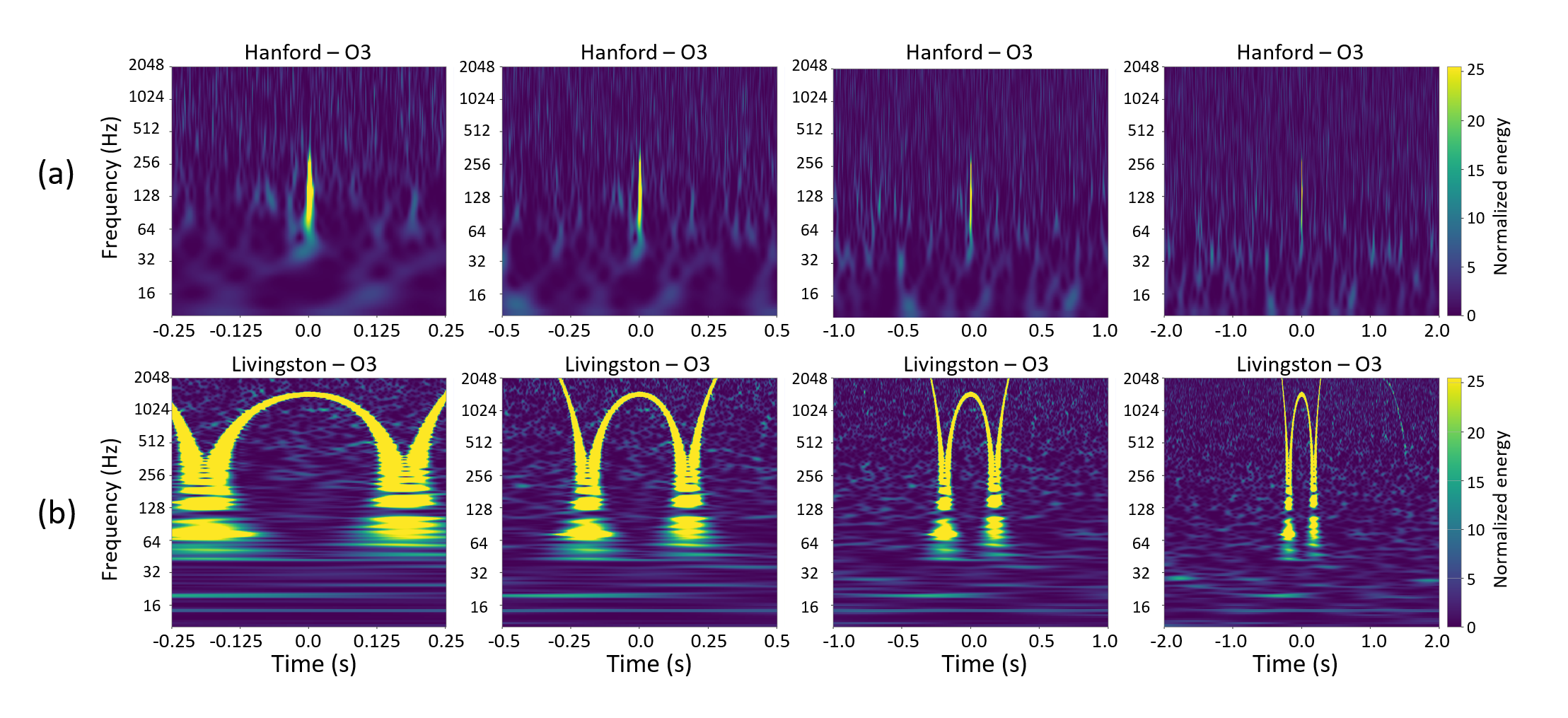}
\end{tabular}
\end{center}
\caption{Time--frequency spectrogram examples of two glitches with four time windows ($0.5~\mathrm{s}$, $1.0~\mathrm{s}$, $2.0~\mathrm{s}$ and $4.0~\mathrm{s}$): 
(a) Blip, characterized by its short duration, and 
(b) Whistle, characterized by its long duration. 
These examples illustrate how different time windows enable the analysis of glitches occurring at various temporal scales, providing valuable insights into the morphological characteristics of each glitch type.}
{\label{fig:glitch_ex}}
\end{figure}

Once data of interest have been identified, a constant-$Q$ transform~\cite{chatterji2004multiresolution, chatterji2005search} is employed to convert whitened LIGO strain data into time--frequency spectrograms, known as Omega scans~\cite{davis2021ligo}. 
Gravity Spy generates four spectrograms for each glitch triggered, using distinct time windows: $0.5~\mathrm{s}$, $1.0~\mathrm{s}$, $2.0~\mathrm{s}$ and $4.0~\mathrm{s}$. 
Each window is centred on the Omicron trigger time.
The incorporation of multiple time windows serves a crucial purpose, allowing both volunteers and machine-learning models to analyze the morphologies of glitches occurring at different characteristic timescales. 
For instance, this enables the examination of short- (e.g., Blip) and long-duration (e.g., Whistle) noise transients, as depicted in Figure~\ref{fig:glitch_ex}.

\begin{figure} 
\begin{center}
\begin{tabular}{c} 
\includegraphics[width=0.90\textwidth]{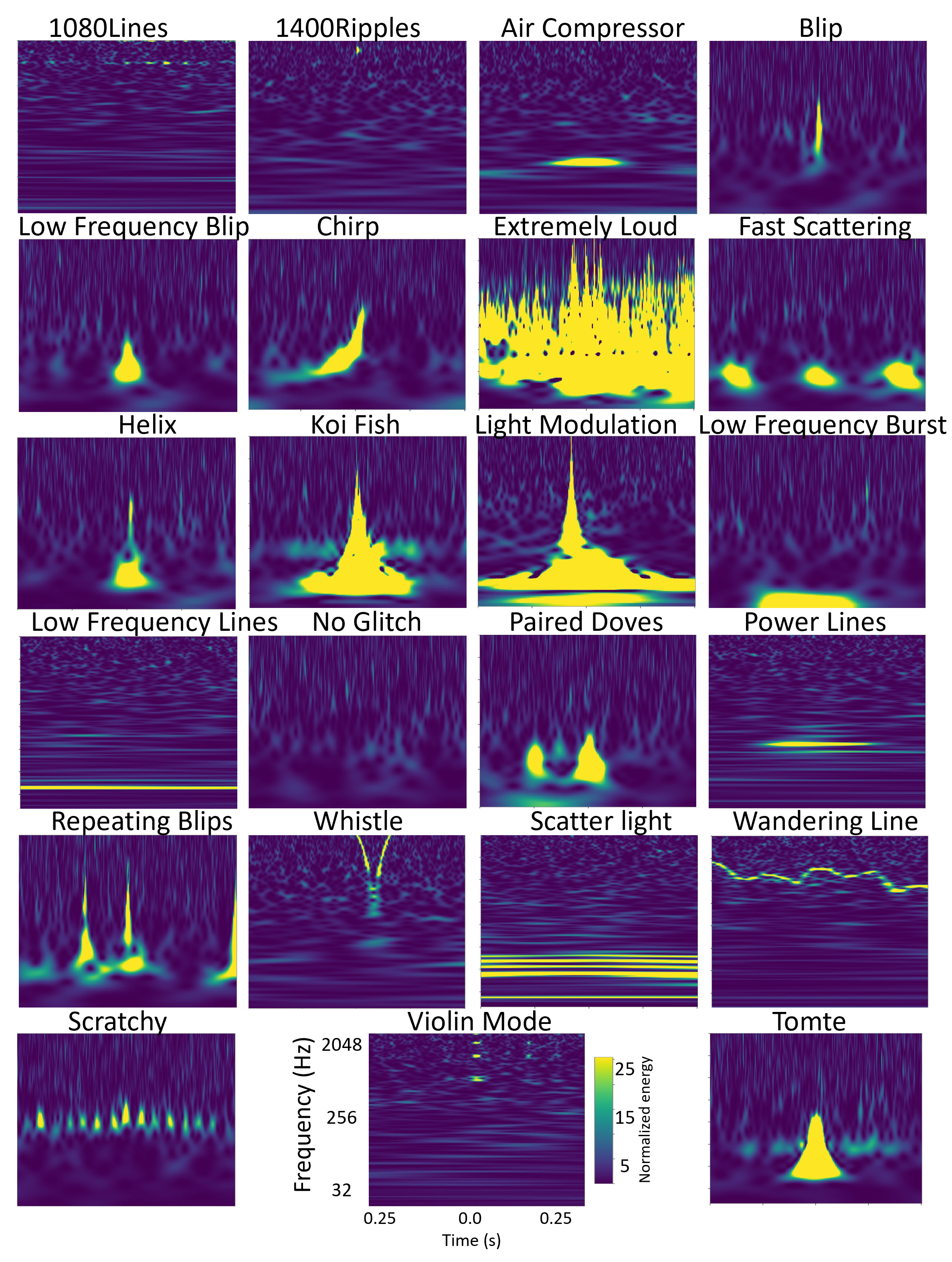}
\end{tabular}
\end{center}
\caption{{Time--frequency morphologies for each of the 23 glitch classes in the Gravity Spy dataset. 
Each image is a spectrogram, with
all sharing the same axes and scale as for the Violin Mode glitch. 
These examples provide one possible morphological representation for each class, while each class can exhibit various shapes and patterns.}}
{\label{fig:all_class}}
\end{figure}

The training and testing datasets employed in this study are sourced from a set of Gravity Spy project classifications~\cite{zevin2022gravity}, focusing exclusively on data from the Hanford and Livingston detectors during O3.  
This classification task has in total 23 glitch classes. 
To ensure consistency, the names associated with each class and the typical morphology of glitches belonging to each class were determined through a collaborative effort between LIGO scientists and citizen scientists. 
Examples representing each of the 23 glitch classes are illustrated in Figure~\ref{fig:all_class}, and more details are provided in the O1--O3 data release paper~\cite{glanzer2023data}.   

For the purposes of this study, we consider the ground truth for glitches in our dataset to be established through a collaborative effort involving volunteers and the previously implemented machine-learning classifier~\cite{zevin2017gravity}. 
However, this ground truth may exhibit a potential noisy-label issue, that is a few glitches were misclassified by the classifier and some volunteers. 
To strengthen the confidence in the ground truth used for our new classification model, we adopt a filtering criterion to create the training and testing dataset. 
Specifically, we included only glitches with consensus confidence scores above $0.9$ from both machine-learning models and volunteer classifications, which accounts for $84.5\%$ of the entire dataset. 
As a consequence, our new model is trained on glitches with relatively reliable labels, effectively reducing confusion during the training process and shielding the model from potentially mislabeled ground truth. Improved training sets for O4 data will be the focus of future studies.

Following the filtering criteria, the Gravity Spy dataset is divided into three distinct sets. 
The training set consists of $8439$ glitches and the validation set consists of $1687$ glitches. 
Additionally, a separate set of $3538$ glitches is reserved for testing purposes. 
Glitches were selected randomly to be partitioned into the three sets.
Each set is carefully curated to ensure no overlapping glitch subjects, thereby preventing data leakage between those three phases. 
A detailed breakdown of the data splitting is presented in Table~\ref{tbl: distribution}. 

\begin{table}[ht]
\centering
\caption{The distribution of each class in the Gravity Spy O3 dataset across the training, validation, and testing sets.}
\label{tbl: distribution}
\begin{tabular}{l r r r}
\hline
 & \multicolumn{3}{c}{Dataset size} \\
 \cline{2-4}
\multicolumn{1}{c}{Gravity Spy class}               & \multicolumn{1}{c}{Train} & \multicolumn{1}{c}{Validation} & \multicolumn{1}{c}{Testing}  \\ \hline \hline
1080 Lines          & 184   & 49         & 225     \\ 
1400 Ripples        & 257   & 64         & 224     \\ 
Air Compressor      & 293   & 57         & 225     \\
Blip                & 312   & 84         & 218     \\ 
Blip Low Frequency  & 337   & 103        & 222     \\ 
Chirp               & 38    & 10         & 17      \\ 
Extremely Loud      & 327   & 85         & 224     \\ 
Fast Scattering     & 515   & 128        & 224     \\ 
Helix               & 46    & 10         & 71      \\ 
Koi Fish            & 338   & 93         & 224     \\ 
Light Modulation    & 127   & 28         & 159     \\ 
Low-frequency Burst & 378   & 106        & 223     \\  
Low-frequency Lines & 381   & 91         & 225     \\
No Glitch           & 311   & 78         & 225     \\ 
Paired Doves        & 124   & 31         & 146     \\
Power Line          & 354   & 83         & 224     \\
Repeating Blips     & 411   & 90         & 224     \\
Scattered Light     & 521   & 128        & 224     \\
Scratchy            & 178   & 36         & 225     \\
Tomte               & 408   & 90         & 225     \\
Violin Mode         & 210   & 56         & 225     \\
Wandering Line      & 25    & 8          & 25      \\
Whistle             & 462   & 129        & 224     \\
\hline 
\end{tabular}
\end{table}

The glitches from all three cohorts undergo the same pre-processing steps. 
Initially, each spectrogram image's original dimension size is $600\times800\times3$ (height, width, color channel). 
We first crop the image with a bounding box defined as [top-left vertical:60, top-left horizontal:100, height:580, width:675] to retain only the intensity region of interest and eliminate irrelevant information, such as axes and scaling maps. 
Next, all cropped glitches are resized to a standardized size of $448\times448\times3$ using bilinear interpolation to facilitate uniformity during model training. 
Finally, min--max normalization is applied to transform each pixel intensity $X$ using ${(X-X_\mathrm{min})}/{(X_\mathrm{max}-X_\mathrm{min})}$, where $X_\mathrm{min}$ and $X_\mathrm{max}$ are the minimal and maximum pixel intensities in the dataset. 
This transformation scales the pixel intensities to a range of $0$ to $1$, ensuring that all images are equally scaled, and enhancing the model's ability to effectively learn from the data.   

\subsection{Multi-view fusion strategy}

As Gravity Spy generates a dataset of glitches with four different durations, it is crucial to have a method for incorporating glitch information of varying characteristic timescales. 
Multi-view fusion, a powerful strategy in machine learning, combines information from multiple sources to enhance the representation and performance of a model. 
Each view contributes unique and complementary insights, leading to improved accuracy, robustness, and generalization of the model~\cite{huang2020fusion}. 
As mentioned above, our Gravity Spy project serves as an example where a single image with four time windows enables models to extract features across diverse timescales. 
Typically, multi-view fusion can employ one of three fusion architectures with different strategies to combine data from various view images. 
We review those below: the performance of these three fusion strategies will be compared in Section~\ref{'sec:3-results'}. 

\begin{figure} 
\begin{center}
\begin{tabular}{c} 
\includegraphics[width=0.95\textwidth]{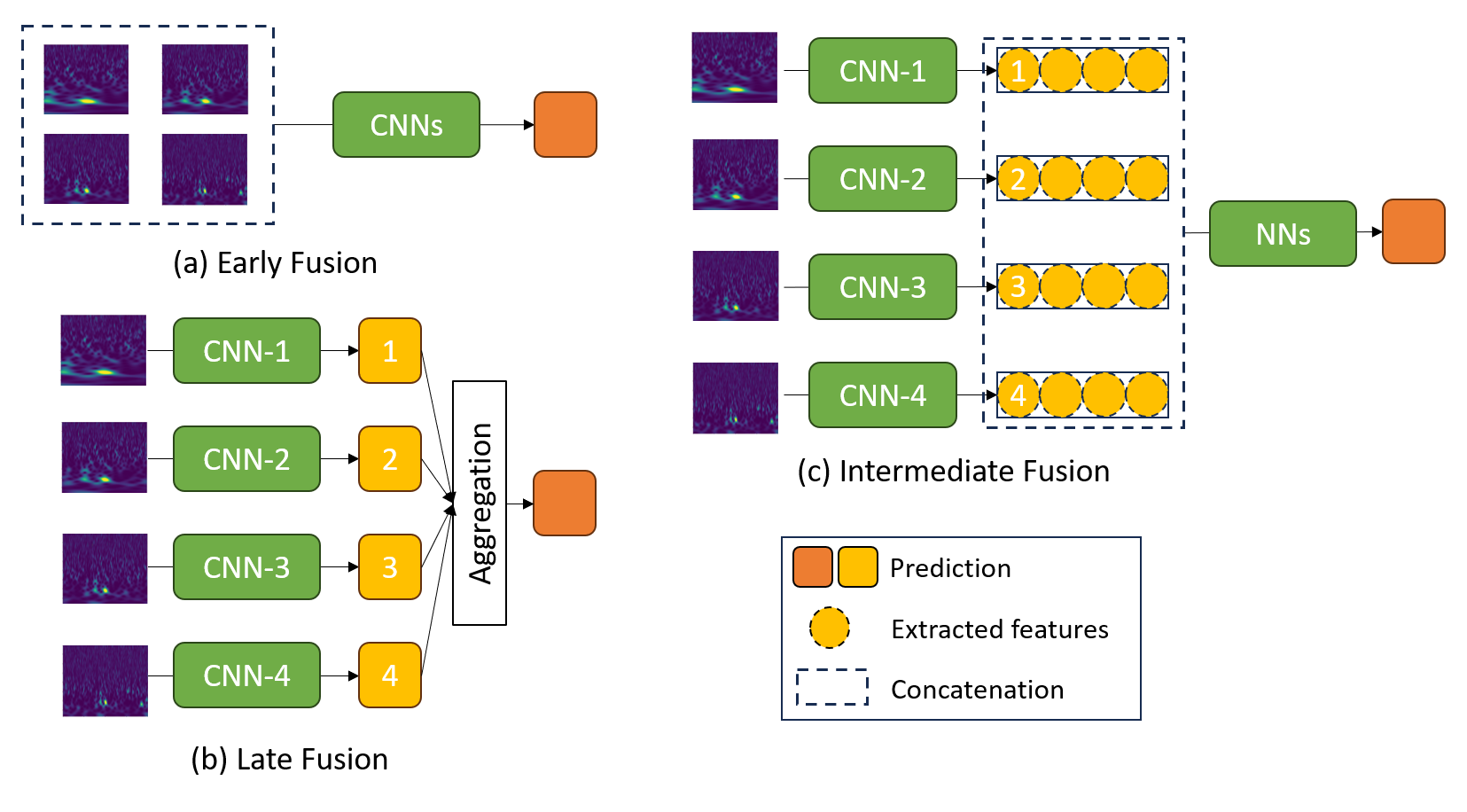}
\end{tabular}
\end{center}
\caption{Illustration of three fusion strategies: (a) early fusion; (b) late fusion, and (c) intermediate fusion. 
CNN: convolutional neural network; NNs: neural networks.}
{\label{fig:fusion}}
\end{figure}

\subsubsection{Early fusion}

Early fusion takes place at the input level, where features from each view are combined in the input layer of the network before being fed into one single model. 
By merging data from multiple views at this stage, the model has access to a more comprehensive representation of the data. 
This approach was employed in previous Gravity Spy classifiers~\cite{soni2021discovering, zevin2017gravity, bahaadini2017deep}, which constructed a $2h \times 2w$ matrix by juxtaposing four $h \times w$ view images, as illustrated in Figure~\ref{fig:fusion}~(a). 

Although it offers several advantages, early fusion has some limitations that may impact its practicality, particularly for large-sized images. 
In previous classifiers~\cite{soni2021discovering, zevin2017gravity, bahaadini2017deep}, the resizing of each view to $140\times170$ was required to fit within the model's limitations, leading to a loss of valuable information. 
Additionally, early fusion assumes equal contributions to the task from all modalities, which may not hold true in this specific project. 
For instance, Repeating Blips~\cite{glanzer2023data} might only appear in the $4~\mathrm{s}$ time-window view and not in others, necessitating the model to assign greater weight to the $4~\mathrm{s}$ window for accurate classification. 

\subsubsection{Late fusion}

Late fusion is a decision-level data-fusion technique, where predictions or outputs from individual models trained on different views are combined at the final stage, as shown in Figure~\ref{fig:fusion}~(b). 
This fusion process often involves aggregation methods, such as majority voting or averaging, to make the final decision. 
The advantage of late fusion is its flexibility in using different models for each view, enabling the selection of the most suitable model for each input.  
However, one drawback of late fusion is its potential to overlook cross-modal relationships, as fusion occurs at a higher decision level rather than at the earlier processing stages~\cite{zadeh2017tensor,humbert-vidan2024comparison}. 
Consequently, in our case, this might lead to suboptimal exploitation of valuable complementarity between different time-window views.

\subsubsection{Intermediate fusion}

Intermediate fusion, also known as feature-level fusion, allows features from each view to be processed separately through different branches of the network, and then combines the resulting feature maps at an intermediate layer, as shown in Figure~\ref{fig:fusion}~(c). 
This technique effectively addresses the limitations of both early fusion and late fusion. 
Specifically, by leveraging the complementarity of different views and capturing cross-modal relationships at a stage where features are still semantically meaningful, intermediate fusion enables the model to learn complex representations and potentially achieve better performance. 

\subsection{Methods}
\label{'model'}

\begin{figure} 
\begin{center}
\begin{tabular}{c} 
\includegraphics[width=0.99\textwidth]{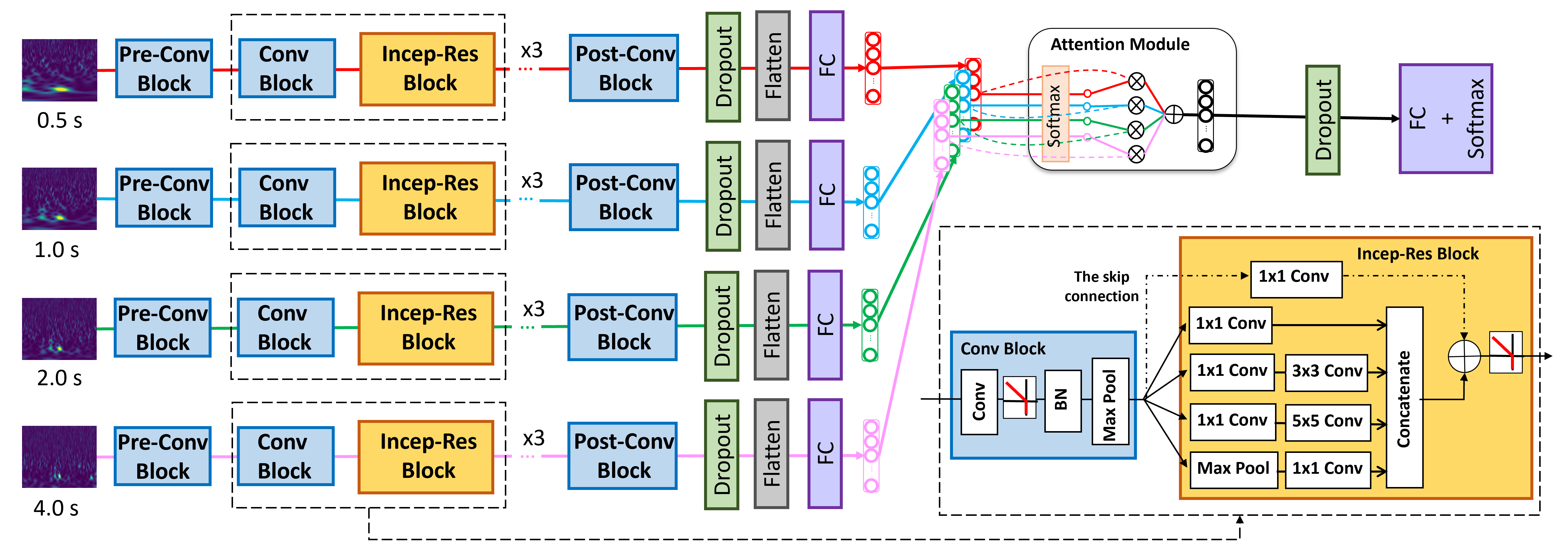}
\end{tabular}
\end{center}
\caption{Architecture of the proposed machine-learning classifier used in O4. Further elaboration on the hyperparameters can be found in Section~\ref{'model'}. 
The $\oplus$ represents element-wise addition, while the $\otimes$ represents multiplication by attention weights in Eq.~(\ref{'eq:attention'}). 
FC: fully connected; BN: batch normalized. }
{\label{fig:model}}
\end{figure}

The proposed classifier distinguishes itself from the past Gravity Spy glitch classifier through several key optimizations, addressing the limitations of previous models~\cite{glanzer2023data, soni2021discovering, zevin2017gravity, bahaadini2017deep}.  
First, it adopts intermediate fusion, effectively combining four views to enhance its representation power by leveraging their complementary information. 
Second, to handle a larger input dimension (higher resolution images), a deeper network is adopted, featuring the inception residual block~\cite{szegedy2017inception} to mitigate the vanishing gradient problem and to extract more discriminative features. 
Additionally, an attention module is introduced at a later stage, providing valuable insights into the model's focus on specific views during decision-making. 
Last, the incorporation of smooth labeling during training effectively addresses the noisy label issue. 
The architecture of the proposed classifier is shown in Figure~\ref{fig:model}, and detailed explanations of these optimizations are presented next. 

\subsubsection{Inception residual block}

The inception residual block combines the ideas of the inception module and the residual block proposed in Szegedy et al.~\cite{szegedy2017inception}. 
As shown in Figure~\ref{fig:model}, it combines these concepts by applying the inception module within a residual block structure. 
Specifically, the block consists of four parallel convolutional branches with filters of different sizes ($1\times1$, $1\times1$, $3\times3$, $5\times5$). 
There is a $1\times1$ convolution, serving as a dimensionality-reduction layer, before the computationally expensive $3\times3$ and $5\times5$ convolutions that reduces the computational cost while preserving the network's capability to learn complex patterns. 
The outputs of the four branches are concatenated, and the resulting feature map is combined with the original input after a $1\times1$ convolutional layer using a skip connection, followed by a rectified linear unit (ReLU) activation function. 
The use of parallel convolutional filters of different sizes allows the block to capture features at multiple scales, improving the model's ability to recognize patterns of varying complexities~\cite{szegedy2015going}. 
This is useful in our case, as different glitch classes have time--frequency features at a variety of scales.
Moreover, the incorporation of skip connections facilitates the training of deep networks, effectively mitigating the vanishing gradient problem~\cite{he2016deep}.

\subsubsection{Attention module}

To assess the varying importance of each view in the final glitch classification, we introduce an attention module $\mathcal{M}_{\mathrm{Att}}$~\cite{ilse2018attention} to the latent vectors of four inputs, denoted as $z_i \in \mathbb{R}^{1\times M}, i=\{1,2,3,4\}$, extracted after the fully-connected layer. 
We set the dimension of the latent vectors to $M = 128$ for this model. 
The attention weight $\alpha_i \in \mathbb{R}, i=\{1,2,3,4\}$ calculated for each latent vector $z_i$ quantifies the significance of each view in making the glitch classification, with all weights summing up to $1$. 
The attention module is defined as
\begin{equation}
  \label{'eq:attention'}  \mathcal{M}_{\mathrm{Att}} = \sum_{i=1}^{4} \alpha_i z_i,
\end{equation}
where the attention weight $\alpha_i$ is computed as
\begin{equation}
    \alpha_i =  \frac{\exp \{ w^\mathrm{T} [ \tanh(Vz_i^\mathrm{T}) \odot \sigma(Uz_i^\mathrm{T}) ] \}}{\sum_{j=1}^{4} \exp \{ w^\mathrm{T} [ \tanh(Vz_j^\mathrm{T}) \odot \sigma(Uz_j^\mathrm{T}) ] \}}.
\end{equation}
The trainable parameters $V \in \mathbb{R}^{L\times M}$ and $w \in \mathbb{R}^{L\times1}$ are associated with the attention mechanism. 
Here, $L$ signifies the one dimension of the matrices for transforming the latent vectors (set to $L = 64$). 
The $V$ matrix is instrumental in identifying similarities among different views. 
The non-linear hyperbolic tangent function $\tanh(\cdot)$ includes both positive and negative values in the gradient flow, but it is almost linear in the $[-1, 1]$ range. 
To handle complex data patterns effectively, we leverage the gated attention mechanism~\cite{ilse2018attention}, which combines $\tanh(\cdot)$ with the element-wise sigmoid function $\sigma(\cdot)$. 
The sigmoid function helps adjust the relevance of information, while the element-wise multiplication $\odot$ facilitates the attention process. 
The gated attention mechanism introduces additional parameters $U \in \mathbb{R}^{L\times M}$ to serve as control gates, mitigating the linearity issue present in $\tanh(\cdot)$. The attention module, widely used in multiple-instance learning tasks~\cite{ilse2018attention, wu2021combining, wu2023smooth, lopez2022deep}, excels in identifying key instances within a bag, i.e., a collection of instances. 

In this paper, we evaluate the impact of the attention module on the model's performance by conducting comparative experiments both with and without the incorporation of attention. 
For the experiments without the attention module, we employed two approaches: averaging the latent vector (i.e., mean pooling) and maximizing the latent vector (i.e., max pooling).

\subsubsection{Label smoothing}

As mentioned in Section~2.1, the ground-truth labels for the Gravity Spy dataset are generated by the previous classifier and volunteers. 
Despite filtering the dataset to only use high-confidence predictions, mislabeled glitches can still be present, potentially confusing the model. To address this problem, label smoothing~\cite{muller2019does}, a valuable regularization technique in machine learning, is applied during model training to enhance generalization and prevent overconfidence in predictions, especially when dealing with noisy or uncertain data. 
This technique introduces a small amount of uncertainty into the training labels. 
Instead of assigning a hard $1$ to the correct class and $0$ to other classes, label smoothing assigns a smoothed value, denoted by $q$, slightly less than $1$ to the correct class while distributing the remaining confidence uniformly across the other classes. 
Mathematically, it can be defined as:
\begin{equation}
\label{eq:smooth}
    q = (1 - \beta)y + \beta u,
\end{equation}
where $y$ represents the true label (encoded as a one-hot vector), $u$ is a uniform distribution over the classes, and $\beta$ is a smoothing parameter between $0$ and $1$. 

In addition, to address the issue of class imbalance (as shown in Table~\ref{tbl: distribution}), we apply different weights $\gamma_i$ to each class $i = \{1,2,\ldots,C\}$ in the loss function during training. 
These weights are inversely proportional to the number of instances in each glitch class and sum up to $1$, ensuring a balanced learning process. 
Therefore, the loss function of the proposed model is defined as 
\begin{equation}
    \mathcal{L}(y, \hat{y}) = -\frac{1}{C} \sum_{i=1}^{C} \gamma_i q_i \log(\hat{y_i}),
\end{equation}
\noindent where $\hat{y}$ represents the predicted probability distribution over classes and $C$ is the number of glitch classes. 
By adjusting the smoothing parameter, the level of smoothing applied to the labels can be controlled. 
For example, setting $\beta=0$ (i.e, no smoothing) leads to one-hot encodings and restores the standard categorical cross-entropy loss, while $\beta=1$ (i.e., completely smooth) results in a uniform distribution over all classes. 

In this paper, the impact of different smoothing levels on the model's performance is also evaluated through different experiments on the value of $\beta$.

\subsubsection{Architecture of the classifier}

The overall architecture of the proposed classifier is shown in Figure~\ref{fig:model}. 
Each of the four branches shares a common architecture. 
Within each branch, a pre-convolutional block is followed by three instances of a combination of a convolutional block and an inception residual block (Incep-Res Block)~\cite{szegedy2017inception}, along with a post-convolutional block.  
These stages effectively extract features from each glitch view. 
Specifically, all convolutional blocks (i.e., Pre-Conv, Conv and Post-Conv blocks) consist of a convolutional layer, a ReLu activation function, and batch normalization (BN) layer and a max pooling (Max Pool) layer. 
Subsequently, a flatten layer and a fully-connected (FC) layer reduce the dimensionality of the feature vectors.
The attention module~\cite{ilse2018attention} then takes these vectors as input, generating a single weighted-averaged feature vector, which is then fed into the final fully-connected layer. 
The softmax activation function yields the probabilities of the $23$ glitch classes. 
To enhance the model's generalization and prevent overfitting, we incorporate an $L2$ regularizer~\cite{cortes2012l2} for each convolutional layer and introduce dropout~\cite{srivastava2014dropout} after each fully-connected layer. 

For optimization, we use the Adam optimizer~\cite{kingma2014adam}, with the learning rate starting at $2 \times 10^{-5}$. 
This choice of optimizer helps to efficiently update the model's parameters during training. 
To address the potential issue of confidence overfitting, where the model may become overconfident in its predictions on the training data, we employ early stopping with a patience of $10$, i.e., if the model does not improve its accuracy on the validation set within $10$ training epochs, we stop training to prevent potential overconfidence.

\subsection{Statistical analysis} 

To ensure the robustness of the model, five independent runs for all experiments are conducted. 
Each of the five runs has different random initializations of the network weights and uses different random mini-batch orderings during training, while keeping the same architecture and hyperparameters. 
For each experiment, we calculate the overall accuracy, precision, F1 score (harmonic mean of the precision and recall), and the area under the receiver operating characteristic curve (AUC) on the held-out test set, where the mean and standard deviation values for each metric are reported. 
To assess the significance of the differences between different model results, we perform a paired $t$-test~\cite{kim2015t} by calculating the differences in scores for each pair, followed by computing the mean and standard deviation of these differences, and determining significance based on a fiducial p-value ($p<0.05$). 
Furthermore, we employ a reliability plot~\cite{brocker2007increasing} to evaluate model calibration across different confidence levels. 
This plot compares predicted probabilities with observed outcomes by binning the probabilities and calculating the average predicted values and observed frequencies for each bin. 
By juxtaposing these elements, the plot reveals the model's accuracy in representing uncertainty. In other words, for a well-calibrated model, if it predicts a set of glitches with a $40\%$ probability of being a Blip, the actual frequency of Blips in that set should also be approximately $40\%$. 
Additionally, calibration accuracy is quantified through the expected calibration error (ECE)~\cite{guo2017calibration}, a statistical metric that measures the average deviation between predicted probabilities and the true likelihood of event occurrence.

%% file: Results/results.tex
\section{Results}
\label{'sec:3-results'}

In this section, we present an overview of our results. We begin by outlining the design of all experiments conducted, providing a comprehensive understanding of our methods. 
Next, we compare the classifier's results under various fusion strategies, evaluate the impact of different smoothing levels, analyze the effects of the gated attention mechanism, and, last, perform a comparative assessment between the proposed model and the previous model implemented in the Gravity Spy pipeline.

\subsection{Experimental design}

In this study, we conduct various ablation studies to assess the model's performance. 
First, the effectiveness of different fusion strategies is compared, as shown in Figure~\ref{fig:fusion}. 
Second, we evaluate different values of the smoothing parameter $\beta$ between $0$ and $1$ with an increment of $0.1$, allowing us to identify an optimal value for $\beta$. 
Third, we compare the proposed classifier with the attention module and a variant without the attention module (i.e., max pooling and mean pooling) to demonstrate the effectiveness of the attention mechanism. 
Fourth, we evaluate the performance of our proposed classifier by comparing it with the previous Gravity Spy model architecture~\cite{bahaadini2018machine, soni2021discovering, glanzer2023data}. 
Both models are trained from scratch using the same training dataset, and their performance is assessed on the same test set. 
Training and testing on the same dataset ensures a fair comparison; these have varied in previous studies (along with the list of glitch classes), so it would not be fair to directly compare statistics to those from other works.
Finally, to better understand the learned feature representations, we employ $t$-distributed stochastic neighbor embedding ($t$-SNE) plots~\cite{van2008visualizing} to reduce the dimensionality of feature vectors obtained after the attention module from $128$ to $2$, offering visualizations for model comparisons in a two-dimensional space. 
Here, for each experiment, we maintain all other experimental conditions at their optimal settings to ensure a fair comparison. 

The experiments and analyses are conducted using TensorFlow~2.11 on two GPUs (NVIDIA Quadro RTX 8000) and scikit-learn, respectively, in Python~3.8. The codes are available via the GitHub of \href{https://github.com/Gravity-Spy/gravityspy-ligo-pipeline}{Gravity Spy}. 

\subsection{Comparison of three fusion strategies}

Table~\ref{tbl: fusion_result} demonstrates the model performance of three fusion strategies on the same test set. 
The proposed intermediate fusion model achieves the best result with an overall accuracy of $0.941$, precision of $0.950$, F1 of $0.931$ and AUC of $0.965$, followed by the late fusion and the early fusion, although the difference between the latter two is not statistically significant. 
Moreover, the results exhibit robustness, as indicated by the minimal standard deviations derived from five independent experiment runs.   

\begin{table}[ht]
\centering
\caption{Model evaluations (the overall mean $\pm$ standard deviation across five independent runs) for three fusion strategies (early fusion, late fusion and intermediate fusion) on the O3 test set. 
Best results are highlighted in bold, and an asterisk (*) indicates that the comparison is statistically significant given our chosen threshold ($p < 0.05$).}
\label{tbl: fusion_result}
\begin{tabular}{lccc}
\hline
 & \multicolumn{3}{c}{Fusion strategy} \\ 
 \cline{2-4}
\multicolumn{1}{c}{Performance metric} & Early fusion & Late fusion & Intermediate fusion\\ 
 & (Figure~\ref{fig:fusion}~(a)) & (Figure~\ref{fig:fusion}~(b)) & (proposed; Figure~\ref{fig:model}) \\  
\hline \hline
Accuracy                                                     & 0.911 ± 0.013                                                 & 0.922 ± 0.022                                                & \textbf{0.941 ± 0.021}*                                                                 \\
Precision                                                    & 0.918 ± 0.008                                                 & 0.920 ± 0.011                                                & \textbf{0.950 ± 0.010}*                                                                 \\
F1                                                           & 0.901 ± 0.009                                                 & 0.901 ± 0.014                                                & \textbf{0.931 ± 0.015}*                                                                 \\
AUC                                                          & 0.939 ± 0.003                                                 & 0.940 ± 0.003                                                & \textbf{0.965 ± 0.004}*                                                                 \\ \hline
\end{tabular}
\end{table}

\subsection{Effects of smoothing labels}

\begin{figure}[ht]
\begin{center}
\begin{tabular}{c} 
\includegraphics[width=0.95\textwidth]{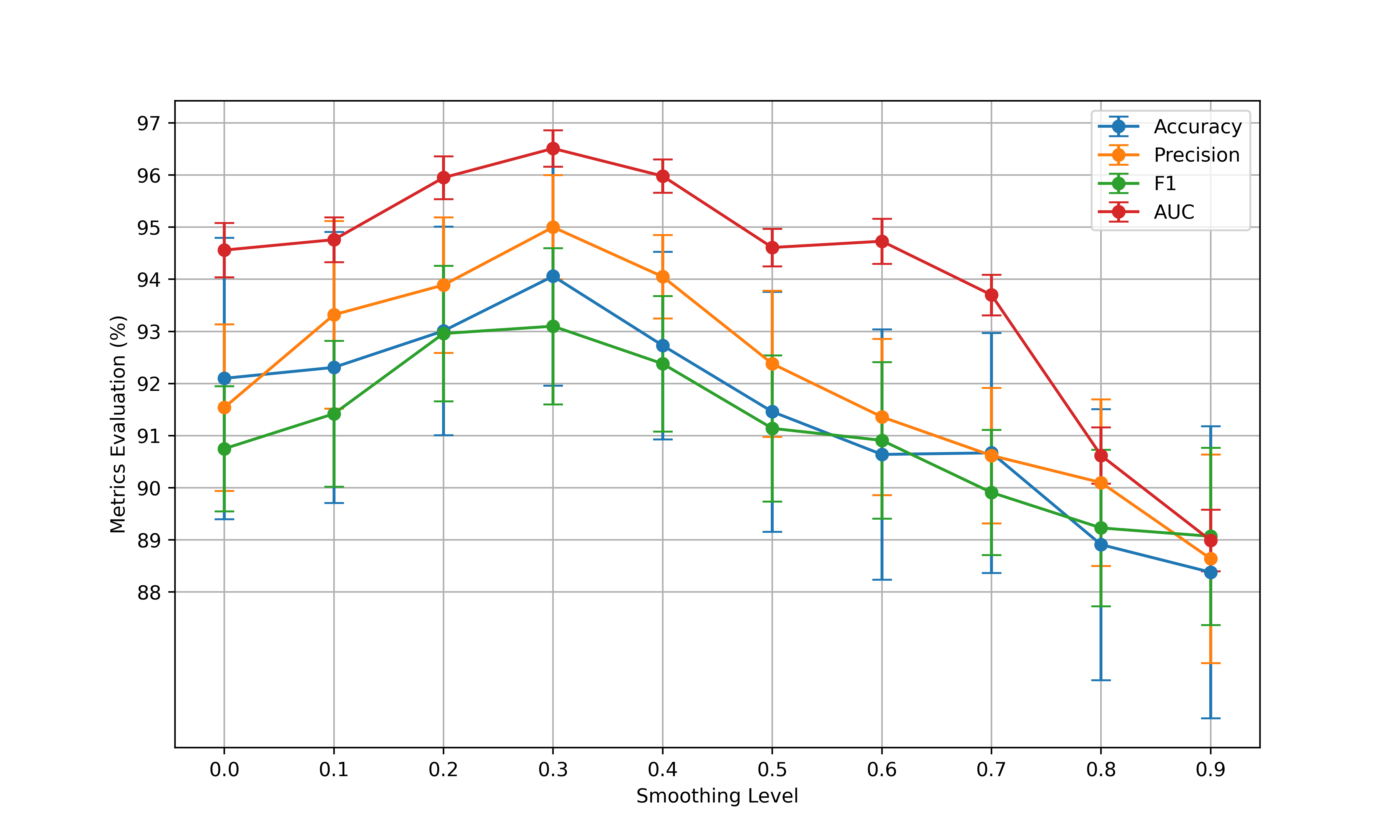}
\end{tabular}
\end{center}
\caption{The evaluation of model performance (accuracy, precision, F1, and AUC scores) on the O3 test set presented across varying smoothing levels ($\beta$) ranging from $0$ to $1$ in increments of $0.1$. 
The results represent the average of five experimental runs, with standard deviations indicated by error bars. 
Here, $\beta = 0$ represents no smoothing.
{\label{fig:smooth_level}}}
\end{figure}

Figure~\ref{fig:smooth_level} presents a comparative analysis of model performance across various smoothing levels of $\beta$ in Eq.~(\ref{eq:smooth}). 
At the outset, for $\beta = 0$, no label smoothing yields a model accuracy of $0.921$, precision of $0.915$, F1 of $0.908$ and AUC of $0.946$. 
With an increase in $\beta$, the model's performance exhibits an initial ascent followed by a subsequent decline. 
When $\beta = 0.3$, the model achieves the optimal performance across all metrics. 
Additionally, the results indicate that, with the implementation of label smoothing, for $\beta \leqslant 0.5$, the model consistently outperforms the one without label smoothing ($\beta = 0$).

\subsection{Effects of gated attention mechanism}

Table~\ref{tbl: att_result} compares the performance of models with and without the gated attention mechanism. 
These models share an identical convolutional neural network architecture for feature extraction from each view but diverge in their feature fusion approaches, such as mean pooling and max pooling. 
The results demonstrate that the performance of the model with the gated attention mechanism is significantly better than of those without it. 
Additionally, statistical analysis indicates no significant difference between the performances of mean pooling and max pooling.

\begin{figure} 
\begin{center}
\begin{tabular}{c} 
\includegraphics[width=0.95\textwidth]{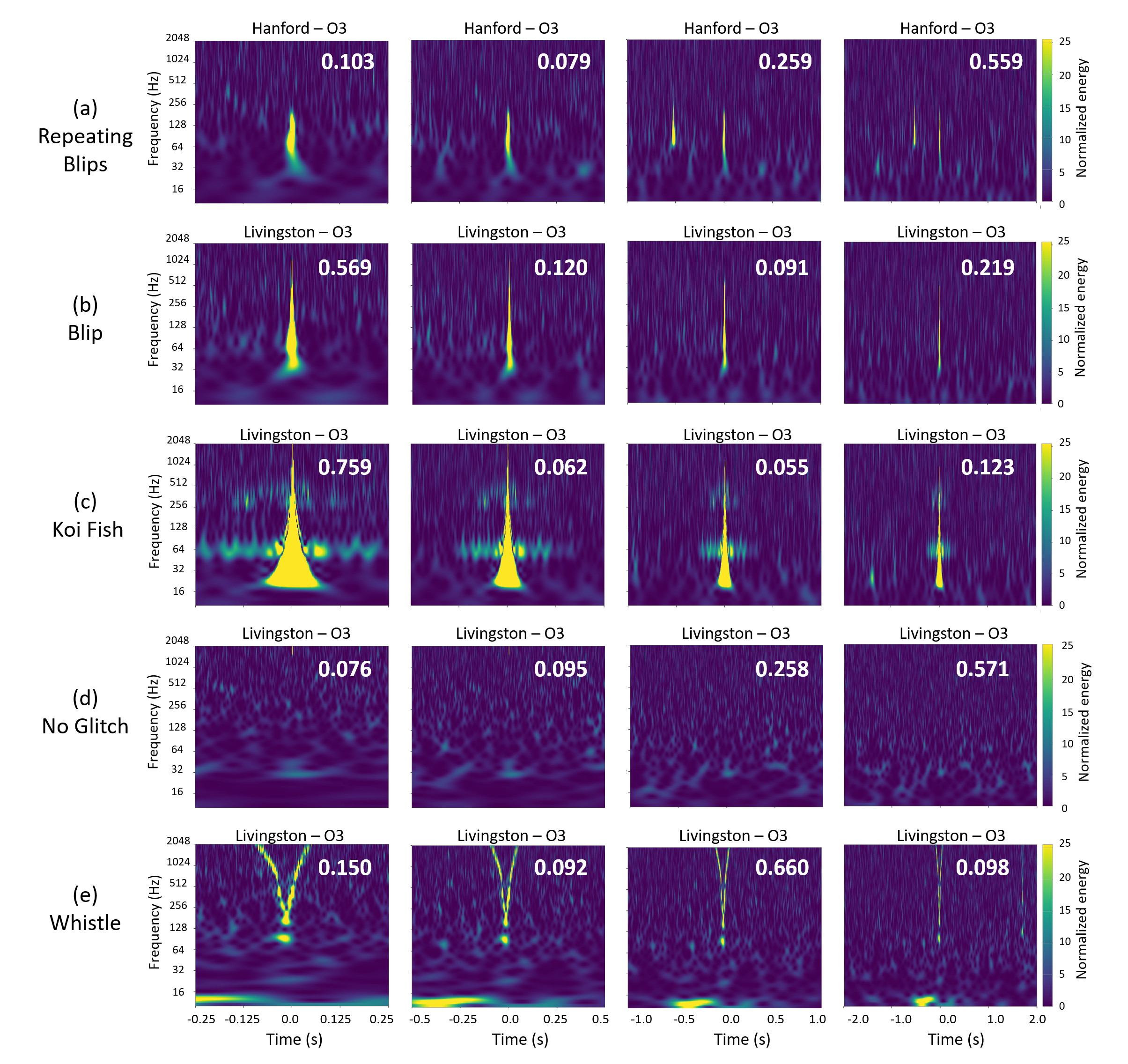}
\end{tabular}
\end{center}
\caption{Illustrative examples of glitch classification with attention weights generated by the classifier for each time-window view. 
Attention weights are shown in white in each subplot. 
A higher attention weight indicates that the model relies more on that particular view for predicting the glitch class. }
{\label{fig:att_ex}}
\end{figure}

The gated attention mechanism also enhances interpretability, rendering the model's outputs more explainable to humans. 
Figure~\ref{fig:att_ex} shows examples with attention weights assigned by the proposed classifier for four glitch predictions. 
The magnitude of attention weights positively correlates with the contribution of each time view to the prediction. 
For instance, in Figure~\ref{fig:att_ex}~(a), (d) and (e), the model places greater emphasis on long time windows to predict glitches like Repeating Blips and No Glitch, while allocating higher attention to short time windows to predict glitches such as Blip and Koi Fish in Figure~\ref{fig:att_ex}~(b) and (c).     

\begin{table}[ht]
\centering
\caption{Model evaluations (the overall mean ± standard deviation across five independent runs) for classifiers with gated attention mechanism and without attention mechanism (using max pooling and mean pooling) on the O3 test set using the intermediate fusion structure. 
Best results are highlighted in bold, and an asterisk (*) indicates that the comparison is statistically significant given our chosen threshold ($p < 0.05$).}
\label{tbl: att_result}
\begin{tabular}{l ccc}
\hline
 & \multicolumn{3}{c}{Attention mechanism} \\ 
 \cline{2-4}
 & Proposed & No Attention & No Attention\\ 
\multicolumn{1}{c}{Performance metric} & (gated attention) & (max pooling) & (mean pooling) \\  
\hline \hline
Accuracy                                                       & \textbf{0.941 ± 0.021}*                                                         & 0.916 ± 0.013                                                          & 0.911 ± 0.011                                                          \\
Precision                                                      & \textbf{0.950 ± 0.010}*                                                         & 0.928 ± 0.010                                                          & 0.929 ± 0.006                                                          \\
F1                                                             & \textbf{0.931 ± 0.015}*                                                         & 0.908 ± 0.011                                                          & 0.903 ± 0.008                                                          \\
AUC                                                            & \textbf{0.965 ± 0.004}*                                                         & 0.948 ± 0.002                                                          & 0.949 ± 0.001                                                          \\ \hline
\end{tabular}
\end{table}

\subsection{Comparison with previous classifiers}

We compare the performance of the proposed model with the previously implemented classifier~\cite{soni2021discovering}. 
The previous classifier has been retrained using the same dataset as our new model to enable a clean comparison.
Results on our O3 test set are shown in Table~\ref{tbl: class_result}. 
The proposed model shows superiority across all metrics, outperforming the previous model by $2.7\%$ to $3.7\%$. 
Furthermore, a detailed examination of class-wise accuracies reveals that the proposed model provides more accurate predictions for the majority of glitch classes. 
In particular, the model performs better for classes like Wandering Line, Repeating Blips, and Helix. 
While there are a few instances for which the previous classifier attains better outcomes, such as Koi Fish and Scattered Light (Slow Scattering), the differences remain relatively modest. 

\begin{figure} 
\begin{center}
\begin{tabular}{c} 
\includegraphics[width=0.95\textwidth]{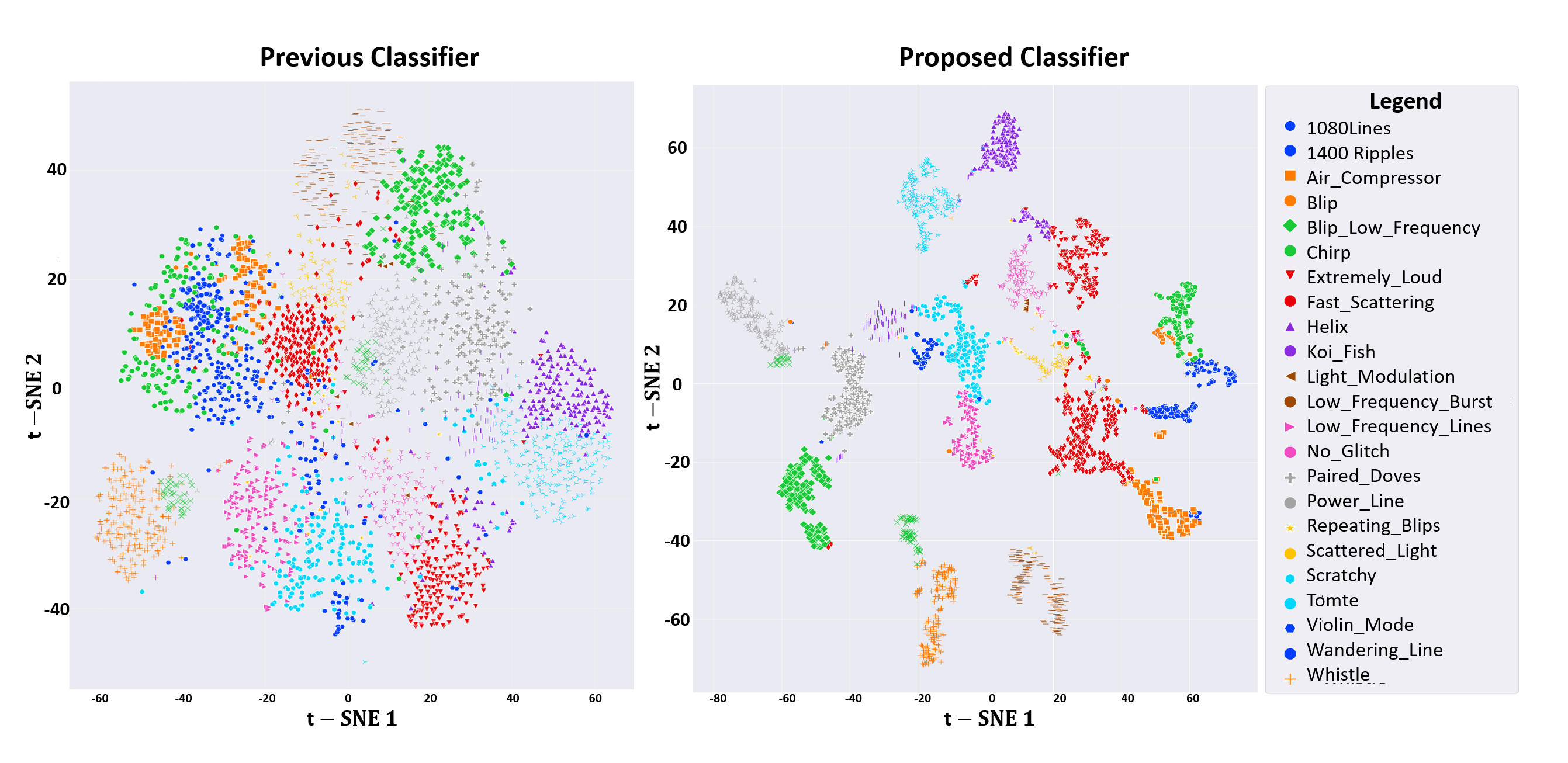}
\end{tabular}
\end{center}
\caption{Comparison of $t$-SNE plots in a two-dimensional space (latent features $t$-SNE~$1$ and $t$-SNE~$2$) between the previous classifier and the newly proposed classifier on the same O3 test set). 
The proposed classifier (right) generates more tightly-clustered glitch features than the previous one (left) within the latent space. 
Both plots were generated using the same $t$-SNE parameters for consistency.}
{\label{fig:tsne_plot}}
\end{figure}

To obtain a deeper insight into this finding, we compare the $t$-SNE plots of these two classifiers within a two-dimensional space, as shown in Figure~\ref{fig:tsne_plot}. 
Compared to the previous classifier, it is evident that the feature distributions generated by the new classifier exhibit tighter clusters within the same glitch class and distinct separation among clusters of different classes. 
However, some degree of overlap persists between glitch classes with similar morphologies, such as Extremely Loud and Koi Fish, Blip and Repeating Blips, Scattered Light and Low-frequency Lines.  

\begin{figure} 
\begin{center}
\begin{tabular}{c} 
\includegraphics[width=0.95\textwidth]{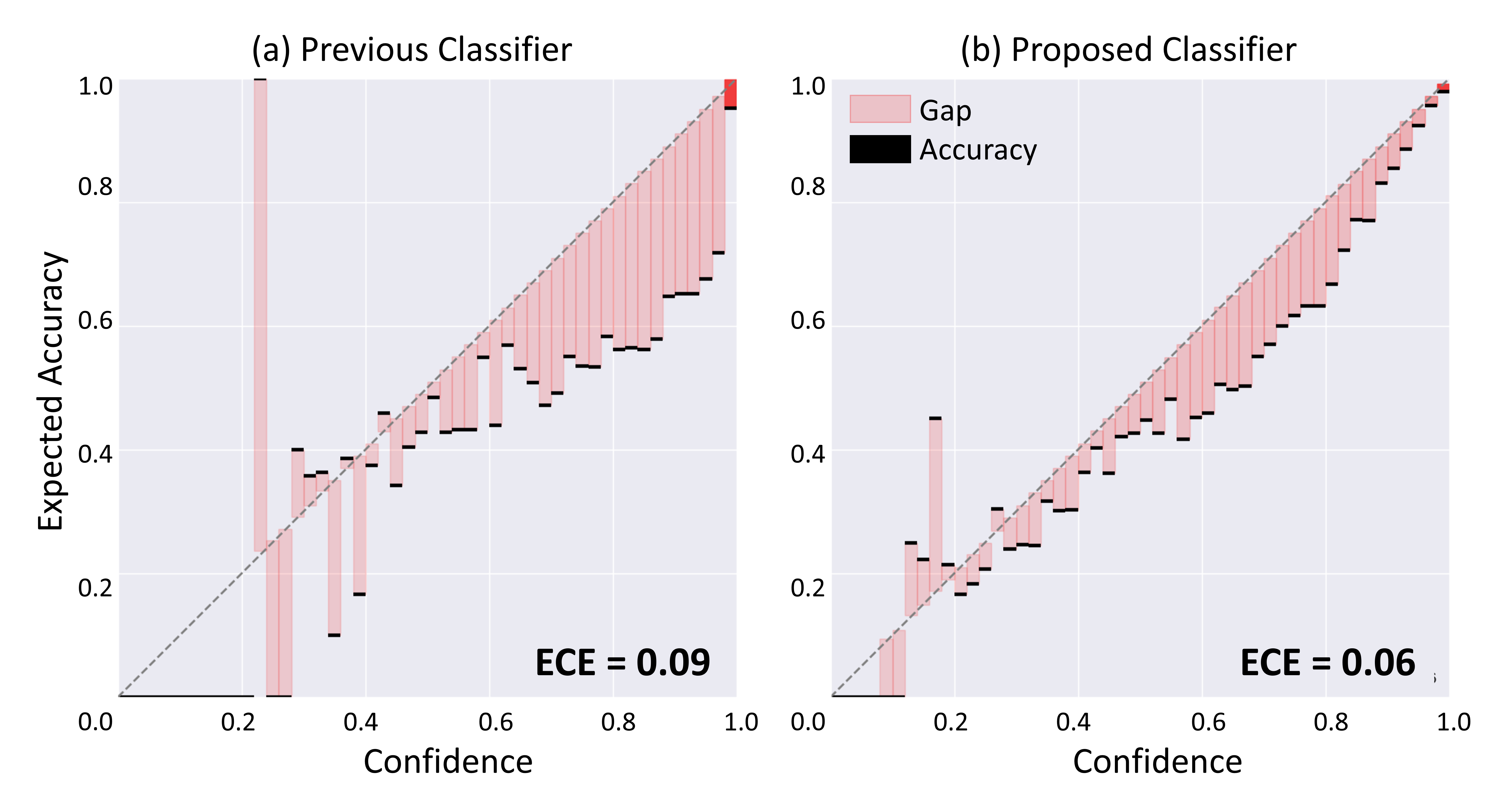}
\end{tabular}
\end{center}
\caption{The reliability plots of the model output probabilities on the O3 test set. 
The plots illustrate the observed fraction of positives against the predicted fraction of positives. 
The diagonal dotted line indicates perfect reliability. The expected calibration error (ECE) quantifies the average discrepancy between predicted probabilities and actual accuracy.}
{\label{fig:ece}}
\end{figure}

Figure~\ref{fig:ece} presents the reliability plots of two classifiers applied to the test set. 
Both classifiers are overconfident in their predictions, evidenced by the majority of points residing below the diagonal line. 
Nevertheless, the proposed model demonstrates better calibration compared to the previous model, particularly in the high confidence range ($> 0.8$)---a crucial aspect for detector characterization and gravitational-wave data analysis. 
Additionally, the proposed model achieves a lower expected calibration error of $0.06$, compared to the previous model's $0.09$. 
This discrepancy indicates that the proposed model displays less bias and higher reliability than the previous classifier.

\begin{table}[ht]
\centering
\caption{Comparison of class-wise classification accuracy, and the overall performance (accuracy, precision, F1 and AUC scores) between the previous classifier and the proposed classifier evaluated with our O3 dataset. 
Better results are highlighted in bold, and an asterisk (*) indicates that the comparison is statistically significant given our chosen threshold ($p < 0.05$).}
\label{tbl: class_result}
\begin{tabular}{l l l l }
\hline
 & & \multicolumn{2}{c}{Classifier} \\
 \cline{3-4}
\multicolumn{2}{c}{Performance metric} & Previous       & Proposed          \\ 
\hline \hline
Class accuracy & 1080 Lines            & 0.985          & \textbf{0.991}    \\ 
               & 1400 Ripples          & 0.967          & \textbf{0.979*}   \\ 
               & Air Compressor        & 0.921          & \textbf{0.954*}   \\ 
               & Blip                  & 0.980          & \textbf{0.991*}   \\ 
               & Blip Low Frequency    & 0.926          & \textbf{0.928}    \\ 
               & Chirp                 & 0.913          & \textbf{0.926*}   \\ 
               & Extremely Loud        & 0.901          & \textbf{0.960*}   \\ 
               & Fast Scattering       & 0.884          & \textbf{0.890}    \\
               & Helix                 & 0.732          & \textbf{0.891*}   \\  
               & Koi Fish              & \textbf{0.921} & 0.917             \\
               & Light Modulation      & 0.918          & \textbf{0.947*}   \\
               & Low-frequency Burst   & 0.962          & \textbf{0.981*}   \\
               & Low-frequency Lines   & 0.818          & \textbf{0.916*}   \\
               & No Glitch             & 0.978          & \textbf{0.984}    \\   
               & Paired Doves          & 0.849          & \textbf{0.882*}   \\
               & Power Line            & 0.987          & \textbf{0.994*}   \\
               & Repeating Blips       & 0.772          & \textbf{0.933*}   \\
               & Scattered Light       & \textbf{0.929} & 0.924             \\
               & Scratchy              & 0.862          & \textbf{0.957*}   \\
               & Tomte                 & 0.963          & \textbf{0.984*}   \\
               & Violin Mode           & 0.924          & \textbf{0.960*}   \\
               & Wandering Line        & 0.520           & \textbf{0.740*}   \\
               & Whistle               & 0.957          & \textbf{0.972*}   \\
\hline
Overall        & Accuracy              & 0.906          & \textbf{0.941*}   \\
               & Precision             & 0.913          & \textbf{0.950*}   \\
               & F1                    & 0.904          & \textbf{0.931*}   \\
               & AUC                   & 0.937          & \textbf{0.965*}   \\ \hline
\end{tabular}
\end{table}

%% file: Discussion/discussion.tex
\section{Discussion}
\label{'sec:4-discussion'}

In this study, we introduce a novel classifier for the Gravity Spy project, aiming to enhance the accuracy of glitch classification provided for LIGO. 
Overall, our proposed classifier achieves accuracy of $0.941$, precision of $0.950$, F1 score of $0.931$ and AUC of $0.965$ on the held-out set of glitches, outperforming prior classifiers integrated into the previous Gravity Spy workflow.   

The proposed classifier exhibits superior glitch classification through various attributes. 
First, the use of the inception residual block empowers the model to deepen its architecture without overfitting concerns, thereby facilitating the extraction of more discriminative features unique to different glitch classes. 
The tightly clustered distributions in Figure~\ref{fig:tsne_plot} further substantiate this finding. 
Next, the intermediate fusion strategy outperforms early and late fusion due to its ability to capture and combine cross-view correlations, which aligns with findings from prior studies, such as using multiple time scales for anomaly detection~\cite{wang2021intermediate} and multiple modalities for gesture recognition~\cite{roitberg2019analysis}. 
Moreover, label smoothing effectively tackles the issue of noisy labels within the training dataset by introducing a controlled level of uncertainty into label distributions during training, resulting in enhanced glitch predictions. 
However, it is crucial to choose the level of this uncertainty infusion appropriately, as excessive smoothing, illustrated in Figure~\ref{fig:smooth_level}, can lead to label confusion within the model, hindering its ability to learn effectively. 
 
The challenge of the black box issue in deep learning stems from the inherent lack of transparency in comprehending how these models arrive at decisions. 
The attention mechanism addresses this challenge by enabling the model to selectively focus on specific features of the input data, assigning varying levels of importance to different regions. 
By highlighting informative features and providing interpretable decisions, the attention mechanism enhances transparency in the decision-making process. 

In our study, the model achieves interpretability by discerning which view contributes the most to the classification decision. 
Consequently, the attention mechanism significantly improves the overall model performance by selectively prioritizing informative aspects of each view and effectively capturing cross-view relationships. 
For example, as shown in Figure~\ref{fig:att_ex}, the model's extraction of more discriminative features from the $2~\mathrm{s}$ and $4~\mathrm{s}$ windows when classifying a glitch as Repeating Blips aligns with volunteers' experiences, reflecting their awareness that repeating patterns primarily manifest in longer time windows. 
In contrast, when identifying a single Blip class, the model focuses on extracting more specific features from the $0.5~\mathrm{s}$ window, corresponding to the transient nature of this particular glitch category. 
Similarly, when dealing with cases resembling Blip or Koi Fish, the model directs greater attention to the $0.5~\mathrm{s}$ window. 
Concerning No Glitch classification, the model's emphasis on the $4~\mathrm{s}$ window is intuitive, aligning with the necessity to inspect the longest window time to confirm all background features and the absence of glitches. 
For Whistle, some attention is given to the $0.5~\mathrm{s}$ window, where the detail of the characteristic up-and-down sweep is most visible, but more is given to the $2~\mathrm{s}$ window. 
While the illustrative Whistle is a single V-shaped sweep, Whistles can often contain multiple sweeps (a W shape as shown in Figure~\ref{fig:glitch_ex} is common), and these are more apparent in the longer windows: the $2~\mathrm{s}$ window seems to give the best balance between resolving the up-and-down structure and the potential repetition of sweeps. 
The emphasis on specific windows does not diminish the importance of those with low attention weights. 
Instead, the model effectively captures crucial information from all four windows, enabling it to make optimal glitch predictions. 
Potentially, the attention mechanism can guide volunteers and LIGO experts, enabling them to enhance their understanding of glitches and thereby facilitating a more effective characterization process.

\begin{figure} 
\begin{center}
\begin{tabular}{c} 
\includegraphics[width=\textwidth]{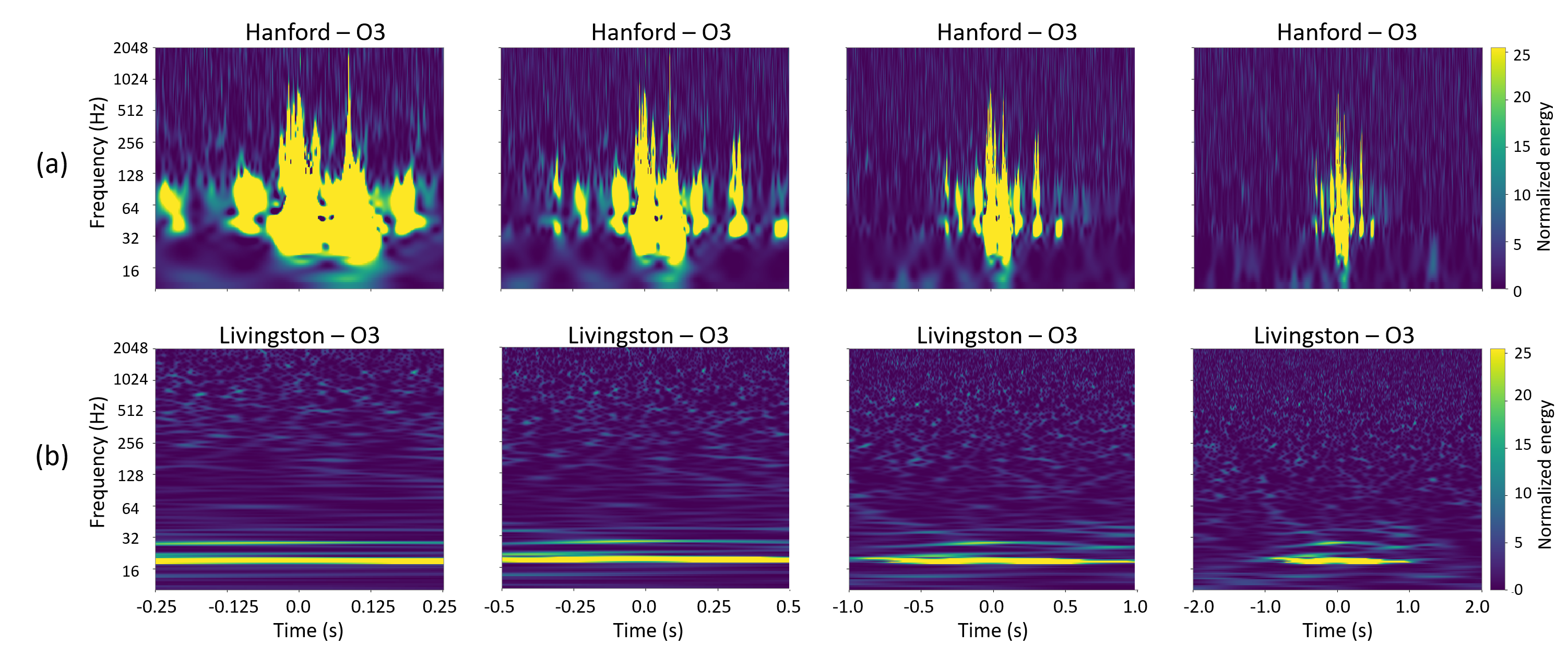}
\end{tabular}
\end{center}
\caption{
Examples of glitches with different predicted classes between the previous model and the proposed model. 
(a) Initially predicted as the Koi Fish class by the previous model, it is now classified as Extremely Loud by the proposed model. 
(b) Initially classified as the Scattered Light class by the previous model, it is now predicted as Low-frequency Lines by the proposed model.}
{\label{fig:mis_plot}}
\end{figure}

Given its superior overall performance in comparison to the previous classifier, the proposed classifier has been integrated into the Gravity Spy system for the O4 run. 
This integration coincides with major infrastructure enhancements introduced in the O4 run~\cite{cahillane2022review, abbott2020prospects, abac2024observation, capote2024advanced}, aimed at achieving higher instrument sensitivity for the detection of gravitational-wave signals. 
However, this improvement in detector sensitivity is accompanied by a trade-off, i.e., a higher sensitivity to environmental and instrumental artifacts, which potentially leads to new glitch classes. 

In this context, the more discriminative features extracted by the proposed classifier, as shown in Figure~\ref{fig:tsne_plot}, have the potential to more effectively identify and categorize these new glitch classes. 
Additionally, Figure~\ref{fig:tsne_plot} also shows distinct clusters within a single class, which suggests the possible existence of sub-classes that may be interesting to explore

An underperformance is observed in certain glitch classes, such as Koi Fish and Scattered Light as presented in Table~\ref{tbl: class_result}. 
However, this underperformance is not significant ($p > 0.05$), approximately $0.5\%$ compared to the previous model, which may potentially be attributed to random sampling variations in our test set. 
Besides, this underperformance is also influenced by the inherent complexity and variability within these classes. Given their prevalence and diverse morphologies, establishing strict boundaries for these classes poses a challenge, leading to expected variations in classifiers when distinguishing these intricacies. 
For instance, in Figure~\ref{fig:mis_plot}, Koi Fish can be confused with Extremely Loud when the trigger is very loud~\cite{soni2021discovering}, and Scattered Light can be confused with Low-frequency Lines as they share characteristics of being low frequency and long duration~\cite{accadia2010noise}.

The previous Gravity Spy classifier has long grappled with the challenge of confidence overfitting. 
This issue can be attributed to various factors, including imbalances in our glitch classes, the presence of noisy labels in our training set, and the use of cross-entropy loss for updating the model parameters. 
These challenges pose the risk of hindering the model's ability to generalize effectively to new and unseen glitch data. 
To address this issue, our proposed classifier incorporates several techniques. 
We apply $L2$ regularization and incorporate smooth labeling into the cross-entropy loss to penalize extreme confidence. 
Moreover, we monitor accuracy for updating model parameters, extending beyond the sole reliance on cross-entropy loss, which tends to drive the predicted confidence close to $1$.

\begin{figure} 
\begin{center}
\begin{tabular}{c} 
\includegraphics[width=0.95\textwidth]{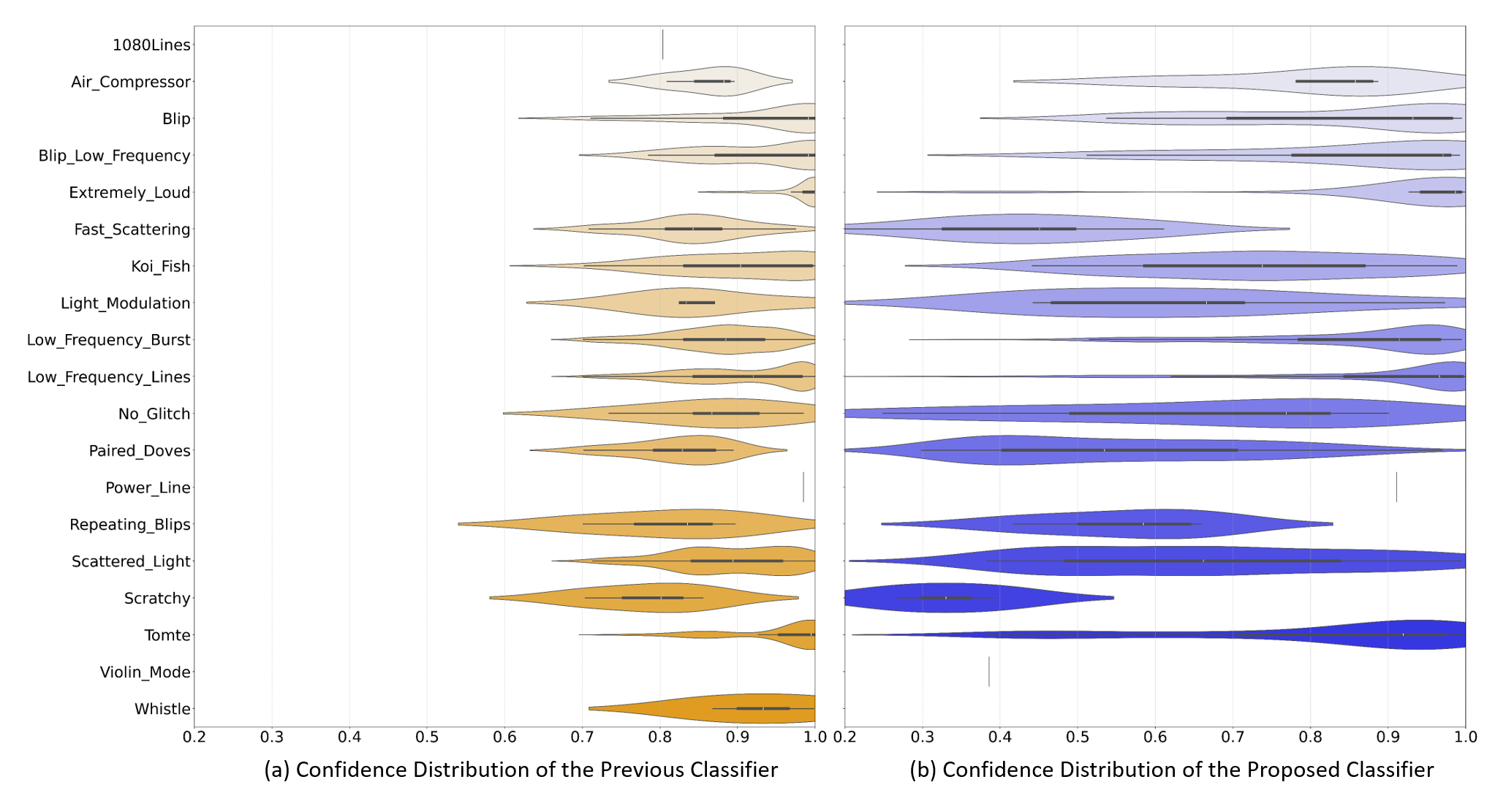}
\end{tabular}
\end{center}
\caption{The density distribution plot of predicted confidence for early O4 glitches between (a) the previous model and (b) the proposed model. The wider sections indicating higher densities. 
The black horizontal bars within each section represent the median confidence score for that glitch class.}
{\label{fig:dist}}
\end{figure}

\begin{figure} 
\begin{center}
\begin{tabular}{c} 
\includegraphics[width=0.95\textwidth]{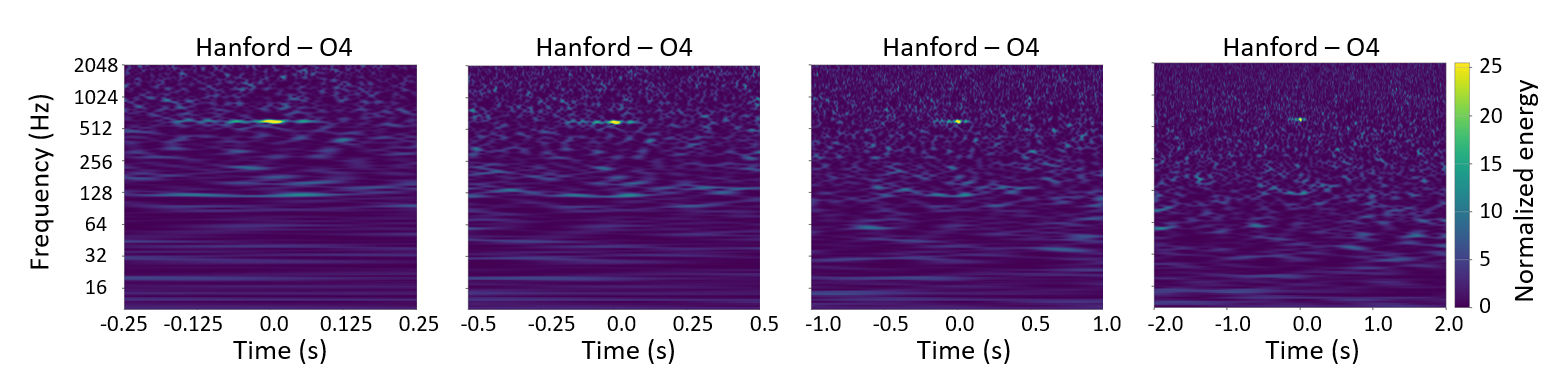}
\end{tabular}
\end{center}
\caption{An example of a new O4 glitch class, named 589 Hz. 
This class appears as a narrow-band line glitch near $589~\mathrm{Hz}$ lasting less than $0.4~\mathrm{s}$, appearing in these spectrograms as bright yellow for $0.02~\mathrm{s}$ to $0.2~\mathrm{s}$. 
This example was classified as the Whistle class with a high confidence of $0.972$ by the previous classifier and a low confidence of $0.348$ by the proposed classifier.  
}
{\label{fig:589hz}}
\end{figure}

To demonstrate the impact of confidence overfitting in O4, we conduct a comparative analysis of the predicted confidence distributions between the previous model and the proposed model. 
This analysis encompasses a total of $2267$ glitches  extracted from early in O4, from 24 May 2023 to 19 September 19 2023 in Hanford and Livingston data. 
As illustrated in Figure~\ref{fig:dist}~(a), the previous model exhibits persistent confidence overfitting, as most glitch classes receive predictions with confidence close to $1$, except for Paired Doves and Light Modulation. 
In contrast, the proposed model, as shown in Figure~\ref{fig:dist}~(b), significantly mitigates this issue, providing predictions with varying confidence ranges across glitch classes.  
Specifically, for glitch classes like Extremely Loud and Low-frequency Lines, which exhibit less variation in morphologies within the O4 distribution compared to O3, the proposed model confidently expresses predictions with relatively high confidence. 
Conversely, for glitch classes such as Whistle, Wandering Line, Violin Mode, and Scratchy, the proposed model displays lower confidence in predictions, given the visually distinct morphologies in O4 compared to their distributions in O3. 
This distinction is crucial for LIGO experts during glitch analysis and characterization, allowing the exclusion of glitches with lower confidences to enhance the precision of their analyses. 
This, in turn, opens avenues to explore the potential discovery of new glitch classes within clusters exhibiting lower confidences. 
A good example is the identification of a new glitch class, named 589 Hz,\footnote{\href{https://www.zooniverse.org/projects/zooniverse/gravity-spy/talk/762/2986261}{www.zooniverse.org/projects/zooniverse/gravity-spy/talk/762/2986261}} depicted in Figure~\ref{fig:589hz}, which occurred on May 30, 31, and June 1 of 2023 in the Hanford detector. 
The previous model predicted it as Whistle with an overall high confidence of $0.976 (\pm 0.059$) on $267$ collections, attributing this classification to its high-frequency nature similar to Whistle, while the proposed model predicted it with significantly lower confidence scores of $0.463 (\pm 0.138$) for $195$ cases classified as Whistle, $0.428 (\pm 0.137)$ for $49$ cases classified as No Glitch and $0.347 (\pm 0.057)$ for $23$ cases classified as Violin Mode, respectively.

This study has some limitations. 
First, as the classifier is trained under full supervision, its generalization capability to new classes is constrained. 
Consequently, if a novel glitch class arises during the O4 run, the classifier requires retraining on a dataset that incorporates the new glitch class. 
This process also involves the participation of volunteers who, with assistance from tools such as the the Gravity Spy similarity search~\cite{coughlin2019classifying}, identify and categorize new glitch classes~\cite{soni2021discovering}. 
Next, the training and test set used in this study are still imperfect with noisy labels. 
To improve the reliability of the model, future efforts should be put into preparing a test set with validated ground truth, as confirmed by LIGO detector-characterization experts. Additionally, there is room for further optimization of the classifier for improved glitch classification. 
Future studies can enhance the model's training by incorporating additional information such as sensor locations, temporal attributes, and supplementary auxiliary data~\cite{zevin2023gravity, nguyen2021environmental, colgan2020efficient, soni2024ligo}. 
For instance, supplementary auxiliary channels, could assist in reducing potential Scattered Light, as illustrated in Figure~\ref{fig:mis_plot}~(b), by aiding in the identification of ground motion \cite{soni2020reducing}. 
Last, our model currently provides information on important views but does not specify which parts of the focused image trigger predictions. 
Our research will explore techniques like Grad-CAM~\cite{8237336} to precisely identify the influential regions within glitch images.

%% file: Conclusion/conclusion.tex
\section{Conclusion}
\label{'sec:5-conclusion'}

In this study, we present a novel classifier for the Gravity Spy project specifically designed to enhance the accuracy of glitch classification for the LIGO gravitational-wave detectors. 
Several factors contribute to the classifier's distinctiveness. 
Leveraging the inception residual block allows for deeper architecture exploration, facilitating the extraction of discriminative glitch features. 
The intermediate fusion strategy captures cross-view correlations more effectively. 
Label smoothing addresses noisy training labels, enhancing the generalization of glitch predictions. 
The attention mechanism not only enhances model performance but also provides interpretability, potentially aiding volunteers and experts in understanding glitch classification decisions. 
Our tests demonstrate that the proposed classifier has exceptional performance compared to the previous classifiers. 

The new classifier has successfully been integrated into analysis of the ongoing O4 run, and is actively employed by both volunteers on the Zooniverse platform and LIGO experts. 
Its capacity to potentially help volunteers identify and categorize new glitch classes aligns with the Gravity Spy project's larger aim of improving detector sensitivity. 
A more advanced model is essential for understanding the details of more intricate glitch patterns. 
Following investigations of new glitch types~\cite{soni2021discovering, soni2024ligo}, the new classifier will be retrained with expanded training data for use in future detector-characterization studies.